\documentclass[epj, twocolumn]{webofc}
\usepackage[varg]{txfonts}  

\woctitle{International Symposium on Very High Energy Cosmic Ray Interactions}

\begin{document}

\title{Composition from high $p_\mathrm{T}$ muons in IceCube}

\author{Dennis Soldin\inst{1}\fnsep\thanks{\email{soldin@uni-wuppertal.de}} for the IceCube Collaboration\thanks{\protect\url{http://icecube.wisc.edu/collaboration/authors/current}}}

\institute{Dept. of Physics, University of Wuppertal, D-42119 Wuppertal, Germany}

\abstract{
Cosmic rays with energies up to $10^{11}\,\mathrm{GeV}$ enter the atmosphere and produce showers of secondary particles. Inside these showers muons with high transverse momentum ($p_\mathrm{T} \gtrsim 2\,\mathrm{GeV}$) are produced from the decay of heavy hadrons, or from high $p_\mathrm{T}$ pions and kaons very early in the shower development. These isolated muons can have large transverse separations from the shower core up to several hundred meters, together with the muon bundle forming a double or triple track signature in IceCube. The separation from the core is a measure of the transverse momentum of the muon's parent particle. Assuming the validity of perturbative quantum chromodynamics (pQCD) the muon lateral distribution depends on the composition of the incident nuclei, thus the composition of high energy cosmic rays can be determined from muon separation measurements. Vice versa these muons can help to understand uncertainties due to phenomenological models as well as test pQCD predictions of high energy interactions involving heavy nuclei.
After introducing the physics scenario of high $p_\mathrm{T}$ muons in kilometer-scale neutrino telescopes we will review results from IceCube in its 59-string configuration as starting point and discuss recent studies on composition using laterally separated muons in the final detector configuration.
}

\maketitle

\section{Introduction}
\label{sec-1}
The primary cosmic ray spectrum has been studied for many decades and incident cosmic rays up to energies of about $10^{11}\,\mathrm{GeV}$ were first observed in the early 1960's \cite{Ref1}. At energies around the knee of the spectrum ($\sim 3$ PeV) direct measurements by balloon and satellite detectors have shown that cosmic rays include nuclei from hydrogen to iron as expected from supernovae \cite{Ref2}. However, at energies above the PeV-range these experiments have very limited statistics and measurements of cosmic rays are only possible by observing extensive air showers with ground-based experiments. In this case studies of the cosmic ray mass composition rely on indirect observables such as the ratio of the measured electromagnetic energy to the number of muons \cite{Ref3, Ref4} or the atmospheric depth $X_\mathrm{max}$, where an air shower reaches the maximum number of particles \cite{Ref5}. These indirect measurements strongly depend on phenomenological models and simulations that relate the observations to the mass composition. Thus, the results will be sensitive to the assumed high energy hadronic interaction models \cite{Ref6}.

High energy ($\gtrsim1\,\mathrm{TeV}$) muons are produced early in the shower development and can therefore be a direct probe of the initial interaction \cite{Ref7}. These muons are produced by the decay of secondary pions and kaons (\emph{conventional muons}) or from the decay of hadrons containing heavy quarks, mostly charm (\emph{prompt muons}) \cite{Ref8}. The bulk of muons produced inside the shower are conventional but at energies above $100\,\mathrm{TeV}$ prompt muons are expected to dominate \cite{Ref9}. 

Inside an air shower high energy muons can be produced that have large transverse momentum ($p_\mathrm{T}$) imparted to them by their parents. These muons will separate from the shower core while traveling to the ground, forming laterally separated muons (LS muons) with separations up to several hundred meters from the core. The resulting lateral separation is a direct measure of the $p_\mathrm{T}$ of the muon's parent. Experimentally a transition from soft to hard interactions, that can be described in the context of perturbative quantum chromodynamics (pQCD), i.e. for values of $p_\mathrm{T}\gtrsim 2\,\mathrm{GeV}$, should be visible in the $p_\mathrm{T}$ spectrum and thereby in the lateral separation distribution. The $p_\mathrm{T}$ spectrum is expected to fall off exponentially with a transition to a power law at roughly $2\,\mathrm{GeV}$ \cite{Ref10}. Moreover, assuming the validity of pQCD calculations the muon $p_\mathrm{T}$ distributions depend on the incident nuclei \cite{Ref111}. Thus, the lateral separation distribution of high energy muons provides a complementary approach to study the cosmic ray mass composition and may help to understand the uncertainties due to phenomenological models as well as test pQCD predictions at high energies and low Bjorken-x.

In IceCube high energy muons are detected with a $1\,\mathrm{km}^3$ array of optical sensors buried at depths between $1450$ and $2450\,\mathrm{m}$ in the Antarctic ice at the South Pole \cite{Ref11}. Using $5160$ digital optical modules (DOMs) on 86 vertical strings Cherenkov radiation produced by charged particles traversing through the ice is detected, such as high energy muons. The surface detector IceTop consists of 162 tanks of ice, each instrumented with two DOMs, that detect cosmic ray air showers at the surface \cite{Ref25}. IceCube and IceTop therefore provide unique opportunities to study high energy muons from cosmic ray air showers.

\begin{figure}[b]
\centering
\includegraphics[width=0.48\textwidth,clip]{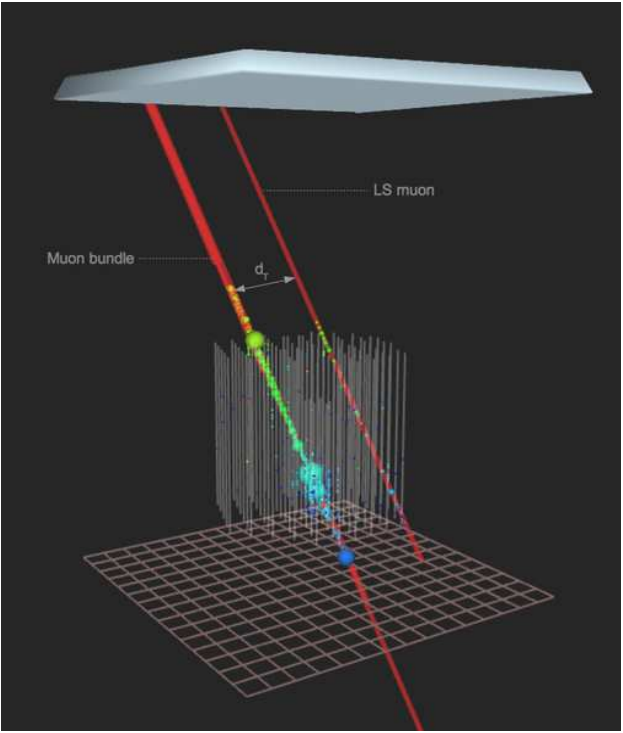}
\caption{Simulated muon double-track event in IceCube. Red lines show the true muon tracks while colored dots represent hits in the detector. Colored spheres around the tracks indicate the light observed by DOMs with time-ordering from red to blue.}
\label{fig-1}      
\end{figure}


\section{Muon lateral separation}
\label{sec-2}
An isolated high $p_\mathrm{T}$ muon together with a muon bundle forming the most compact region of the shower, produce a \emph{double track signature} in IceCube. Figure \ref{fig-1} shows a typical simulated double track event in IceCube. The muon's lateral separation from the shower core is related to the muon's $p_\mathrm{T}$ via
\begin{equation}
\label{eq-1}
d_\mathrm{T}\simeq \frac{p_\mathrm{T}\cdot H_\mathrm{int}}{E_\mu\cdot \cos(\theta)},
\end{equation}
where $H_\mathrm{int}$ is the primary interaction height, $\theta$ is the zenith angle and $E_\mu$ is the muon's energy at generation which is well approximated by the energy at the Earth's surface. For the relevant muon energies and separations the effect of Earth's magnetic field and multiple scattering is negligible. Hence, the initial $p_T$ is the dominant effect producing the separation, although the separation distribution will be biased toward events produced at high altitudes, high $p_\mathrm{T}$, and low energy \cite{Ref10}. The horizontal spacing between the IceCube strings of $125$ m will be a rough estimate of the minimum resolvable track separation. For a $1\,\mathrm{TeV}$ vertical muon produced at an altitude of $50\,\mathrm{km}$ this corresponds to a $p_\mathrm{T}$ of about $2.5\,\mathrm{GeV}$. More inclined arrival directions will decrease this threshold. Thus in this work we will consider transverse momenta of $p_\mathrm{T}\geq 2\,\mathrm{GeV}$.

In high energy primary interactions the parents of LS muons will most likely be produced in high $p_\mathrm{T}$ particle jets. Due to energy and momentum conservation high $p_\mathrm{T}$ jets are typically produced back-to-back in the center of mass frame forming a dijet event \cite{Ref12}. If in both jets high $p_\mathrm{T}$ particles decay into muons, two isolated muons and the central muon bundle can produce a \emph{triple track signature} in IceCube where the two isolated muons are located roughly on opposite sites of the bundle. Although the expected flux will be low, this would be a distinctive signature and a direct probe of hadronic high energy cosmic ray interactions.

A first analysis of laterally separated muons in IceCube was performed using one year of data in the 59-string configuration \cite{Ref10}. In this analysis high energy data is used where events are required to generate at least 630 photoelectrons corresponding to a primary energy threshold of about 1 TeV \cite{Ref13}. After high energy filtering $6.4 \times 10^7$ events remain in the data sample and additional reconstructions are applied.

\subsection{Reconstruction}
\label{sec-21}
Double track signatures have a unique topology: a muon bundle accompanied by a LS muon. Thus a dedicated reconstruction algorithm based on a two track hypothesis is needed to reconstruct the muon bundle and a LS muon with the same timing and direction.
A fast first guess reconstruction is used that is based on a linear relationship between arrival times of the Cherenkov light at the DOMs and the wavefront which gives the approximate direction of the muon bundle. The hits are rotated into a plane perpendicular to the first guess track and sorted into two sets of hits using the \emph{k-means clustering algorithm} \cite{Ref14} that sorts the hits according to the closeness of the mean of the cluster. The larger cluster of hits is expected to belong to the muon bundle and is reconstructed using a maximum-likelihood function that accounts for the arrival time of the Cherenkov photons and the scattering of light in ice \cite{Ref15}.
Hits that belong to the LS muon are identified by their arrival time relative to the muon bundle: only hits from the smaller cluster that arrive $100\,\mathrm{ns}$ earlier than the expectation for light from the muon bundle are considered as LS muon hits. Afterwards this method is iterated to improve the accuracy of the reconstruction, especially to eliminate the few cases where the initial clustering algorithm didn't perform well. The LS muon hits are then reconstructed seperately using the previous maximum likelihood function.  

\subsection{Event selection}
\label{sec-22}
Only events are kept where both the reconstruction of muon bundle track as well as the LS muon track reconstruction succeeded, i.e. the likelihood fits found a minimum. The main background of this analysis are cosmic ray air showers without a laterally separated muon that can be divided into two distinct types of background: single air showers and multiple independent coincident showers.

Background from multiple coincident showers is reduced by requiring that the LS muon and muon bundle track directions agree within $5^\circ$ and that they arrive within $\pm 450$ ns from each other. An irreducible background remains from showers that arrive from the same direction and at the same time. This background is estimated with  \emph{off-time} data, events from the same direction but arriving between $450$ and $1350\,\mathrm{ns}$ from each other.

Single shower events can be eliminated using the topological properties of single and double tracks. Therefore a reduced likelihood function made for a single track hypothesis is used where single showers are well-reconstructed while double tracks are not. Requiring an output of the (negative) reduced likelihood larger than $7.5$ reduces the single shower background significantly. The closer two tracks are, the more difficult it is to separate them. Hence, events with track separations $d_\mathrm{T}<135\,\mathrm{m}$ are removed. To remove surviving events that still show poorly reconstructed directions, we require LS muon hits on more than two strings and the LS muon track to trigger more than 8 DOMs. 

\begin{figure}[b]
\centering
\includegraphics[width=0.48\textwidth,clip]{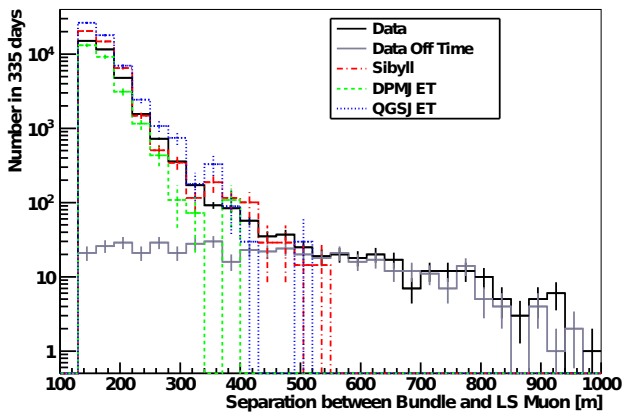}
\caption{The lateral distribution of muons after applying all selection criteria for IC59 data and simulations, as well as irreducible background estimated from off-time data \cite{Ref10}.\newline}
\label{fig-2}      
\end{figure}

\subsection{Results}
\label{sec-23}
After applying all selection criteria $34,754$ events remain in the data, where the expected number of random coincident showers obtained from off-time data is $456$. Figure \ref{fig-2} shows the lateral separation distribution of the remaining events as well as the irreducible background estimated from off-time data. Also shown are model predictions from Sibyll 2.1 \cite{Ref16}, QGSJET01c \cite{Ref17}, and DPMJET 2.55 \cite{Ref18} generated using the CORSIKA \cite{Ref19} simulation program where the simulated cosmic ray spectrum is based on the H\"orandel polygonato model \cite{Ref20} with primary energies between $600\,\mathrm{GeV}$ and $10^{11}\,\mathrm{GeV}$.

\begin{figure}[t]
\centering
\includegraphics[width=0.48\textwidth,clip]{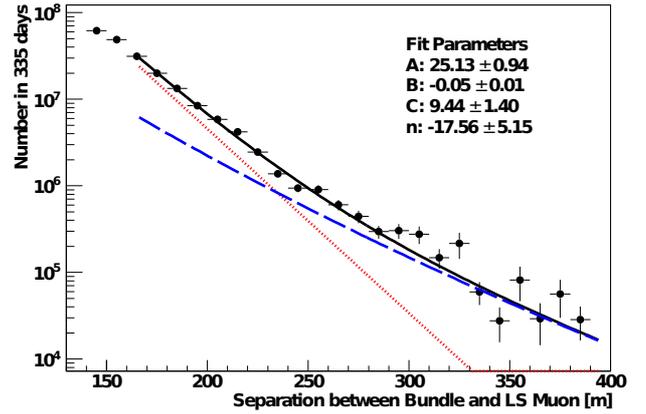}
\caption{The lateral data distribution at sea level and the best fit parameters for equation (\ref{eq-2}). The exponential part is plotted as dotted red line and the power law as dashed blue line \cite{Ref10}.}
\label{fig-3}      
\end{figure}

To test the expected transition from an exponential to a power law in the lateral distribution the distribution was translated to sea level after subtracting the background and then fitted with an exponential function that transitions into a power law
\begin{equation}
\label{eq-2}
N=\exp(A+Bx)+10^C\left(1+\frac{x}{400}\right)^n
\end{equation}
with A, B, C, and n allowed to vary. The resulting fit and the corresponding best fit parameters are shown in Figure \ref{fig-3}. The fit has a $\chi^2/\mathrm{DOF}$ of $30.8/19$ with a probability of $4\%$ being a good fit for this distribution while a fit to a purely exponential function has a $\chi^2/\mathrm{DOF}$ of $60.5/21$ with a probability of $0.001\%$. Hence, an exponential that transitions to a power law is favoured, as expected from pQCD.

\begin{figure}[b]
\centering
\includegraphics[width=0.48\textwidth,clip]{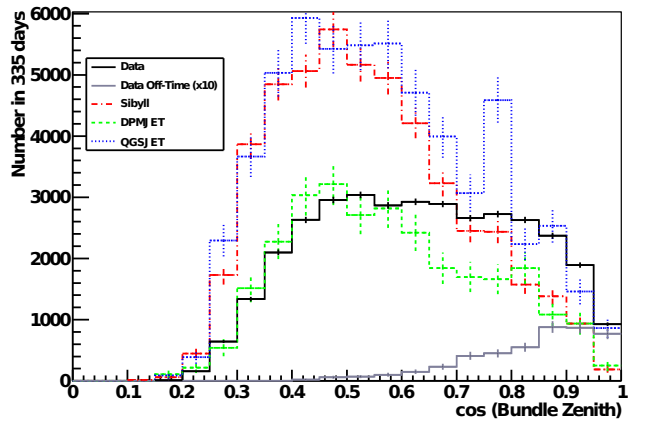}
\caption{Zenith angle distribution of the reconstructed muon bundle track after applying all selection criteria for IC59 data and simulations, as well as irreducible background estimated from off-time data \cite{Ref10}.}
\label{fig-4}      
\end{figure}

While different hadronic interaction models seem to describe the track separation distribution of the experimental data fairly well, the arrival angle distribution shows a significant disagreement between simulation and data. Figure \ref{fig-4} shows the cosine of the zenith angle of the muon bundle for events that survive all selection criteria. While Sibyll and QGSJET overpredict the event rate at high zenith angles and underpredict the rate for more vertical events, DPMJET seems to describe the data better, at least at high zenith angles. A Kolmogorov-Smirnov test of the simulated distributions against the data finds that none of the simulations are in good agreement. The highest correlation is found for DPMJET with a probability of only $5\times 10^{-12}$ that it has been drawn from the same distribution as the data \cite{Ref10}.

Besides different approaches of $p_\mathrm{T}$ modeling between the interaction models it is interesting to note that DPMJET is the only model that includes a hard component of heavy hadrons including charm while agreeing best with data. However, the zenith angle distribution of prompt muons for DPMJET is rather flat, thus the impact on the zenith angle discrepancy is expected to be small. Moreover, further studies have shown that the difference in zenith angle could arise from a larger fraction of muons from kaon decays. DPMJET for example shows a higher fraction of muons produced by kaons at high zenith angles compared to QGSJET \cite{Ref10}. The discrepancy may also be related to the primary mass composition, since the interaction models show a different behavior in zenith angle for different primary compositions. This indicates that an interplay between kaon/pion treatment, charm abundance, and composition could explain the zenith angle discrepancy.

\section{Composition from high $p_\mathrm{T}$ muons}
\label{sec-3}

pQCD predictions of $p_\mathrm{T}$ spectra depend on the mass composition of the initial cosmic rays \cite{Ref111}. Nuclei with energy $E_{\mathrm{prim}}$ and atomic number $A$ have the same parton distributions as $A$ nucleons, each with energy $E_{\mathrm{prim}}$ divided by $A$. Since the lateral separation of muons is a direct measure of the initial $p_\mathrm{T}$ it will also be sensitive to the cosmic ray composition. As pointed out, the zenith angle discrepancy, for example, may be related to differences in mass composition. A Kolmogorov-Smirnov test indicates that proton-only distributions show the best match with experimental data from IC59 but no final conclusions have been drawn.

\begin{figure}[t]
\centering
\includegraphics[width=0.48\textwidth,clip]{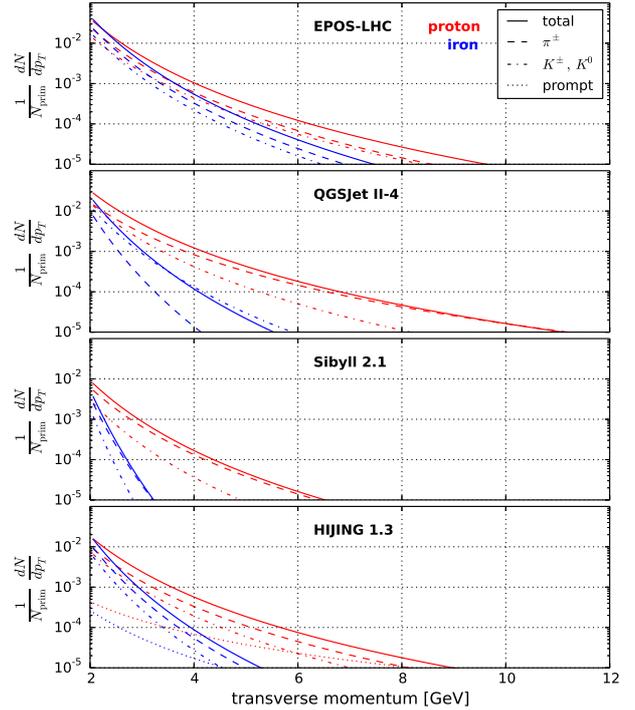}
\caption{Simulated $p_\mathrm{T}$ distributions for $E_{\mathrm{prim}}=10^5\,\mathrm{GeV}$ and different hadronic interaction models for proton (red) and iron (blue) primaries normalized to the number of collisions $N_\mathrm{prim}$.}
\label{fig-5}      
\end{figure}

\begin{figure}[t]
\centering
\includegraphics[width=0.48\textwidth,clip]{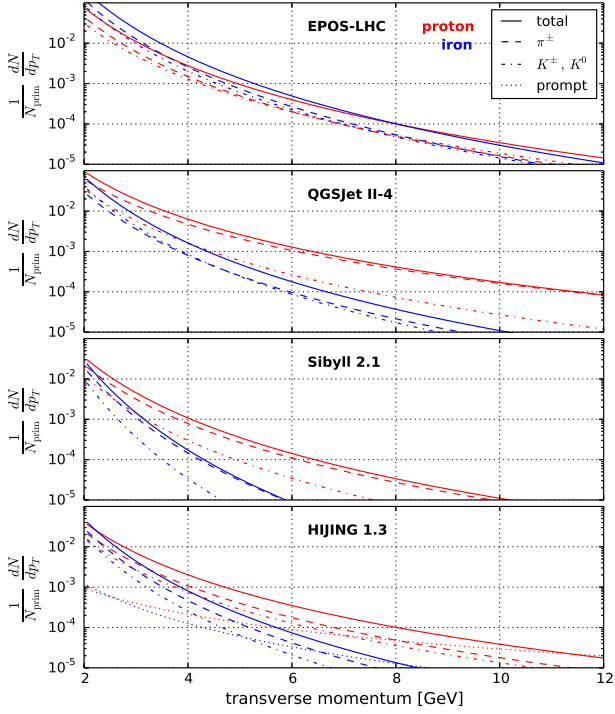}
\caption{Simulated $p_\mathrm{T}$ distributions for $E_{\mathrm{prim}}=10^6\,\mathrm{GeV}$ and different hadronic interaction models for proton (red) and iron (blue) primaries normalized to the number of collisions $N_\mathrm{prim}$.}
\label{fig-6}      
\end{figure}

To improve the situation, more sophisticated studies on the influence of primary mass composition on high $p_\mathrm{T}$ muons are pursued. Therefore the first interaction of the cosmic ray with the atmosphere\footnote{Assuming nitrogen as target element as it forms approximately $78\%$ of the atmosphere and has approximately the same mass number as oxygen which forms roughly $21\%$ of the atmosphere.} for different primary energies and mass compositions is simulated using the CRMC package \cite{Ref21} as interface to get access to different hadronic interaction models. Generating $5\times 10^4$ collisions for each initial energy of $\log_{10}(E_{\mathrm{prim}}/\mathrm{GeV})=4.0,\,4.1,\,\dots,\,7.9,\,8.0$ allows obtaining transverse momentum and energy spectra of particles that may decay into muons: pions ($\pi^\pm$), kaons ($K^\pm$, $K^0$), and heavy hadrons (prompt: $D^0$, $D^+$, $D_s^+$, $\Lambda_c^+$, $\Omega_c^0$, $B^0$, $B^+$, $B_s^+$, $B_c^+$, $\Lambda_b^0$, $\Sigma_b^0$, $\Sigma_b^+$ and their antiparticles). The resulting $p_\mathrm{T}$ spectra are fitted in the range $2\,\mathrm{GeV}\leq p_\mathrm{T}\leq 12\,\mathrm{GeV}$ with a power law of the form
\begin{equation}
\label{eq-3}
N(p_\mathrm{T})=a\cdot p_\mathrm{T}^{b},
\end{equation}
where $a$ and $b$ are allowed to vary. For a given primary energy only a weak correlation between the secondary particles's $p_\mathrm{T}$ and its energy is assumed and the energy spectra of each component can also be obtained from the simulations. Figures \ref{fig-5} and \ref{fig-6} show the resulting $p_\mathrm{T}$ spectrum fits obtained from EPOS-LHC \cite{Ref22}, QGSJET II-4 \cite{Ref23}, Sibyll 2.1 \cite{Ref16}, and HIJING 1.3 \cite{Ref24} for primary energies of $10^5\,\mathrm{GeV}$ and $10^6\,\mathrm{GeV}$ respectively, which is the typical energy range of cosmic ray events in IceCube after high energy filtering. The distributions are shown for different primary mass compositions: proton-only (red) and iron-only (blue). Note that HIJING is the only hadronic interaction model that includes a prompt component. Although different interaction models give very different $p_\mathrm{T}$ spectra the iron distributions show a significantly steeper spectrum compared to the proton spectra for all models and energies considered here, as expected. 

It is interesting to note that especially in iron collisions EPOS produces a significantly larger number of high $p_\mathrm{T}$ particles per collision with respect to other models while Sibyll seems to underestimate the number of high $p_\mathrm{T}$ pions and kaons. Moreover, different interaction models produce very different fractions of high $p_\mathrm{T}$ pions and kaons: Sibyll seems to generate the lowest number of kaons with high transverse momentum compared to other models while EPOS and QGSJET show the largest fraction of kaons. As pointed out before, this may affect the zenith angle distributions and could explain the observed zenith angle discrepancy in IC59. Also notice that the prompt spectra obtained from HIJING show a significantly flatter slope compared to the pion and kaon contributions, as expected from pQCD. Hence, studies of high $p_\mathrm{T}$ muons seem to be also sensitive to the prompt muon flux, especially at energies where prompt muons are expected to dominate ($\gtrsim 100\,\mathrm{TeV}$).

\begin{figure}[b]
\centering
\includegraphics[width=0.48\textwidth,clip]{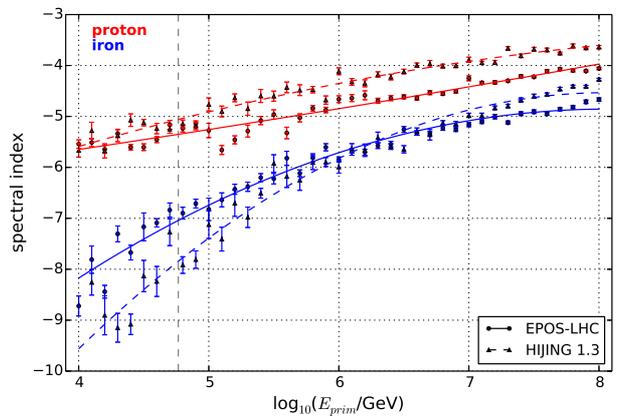}
\caption{Spectral indices $b$ obtained from power law fits of the form $N(p_\mathrm{T})=a\cdot p_\mathrm{T}^{b}$ applied to EPOS and HIJING simulations for proton (red) and iron (blue) primaries respectively and corresponding spline fits (lines). Assuming an $E^{-\gamma}$ primary spectrum $95\%$ of the events are generated with initial energies $\log_{10}(E_{\mathrm{prim}}/\mathrm{GeV})<4.77$ (dashed gray line).}
\label{fig-7}      
\end{figure}

To obtain $p_\mathrm{T}$ and energy spectra of secondary pions, kaons, and the prompt component for a given primary energy, each generated $p_\mathrm{T}$ spectrum is fitted with a power law of the form (\ref{eq-3}) and a spline fit is applied to the resulting spectral indices $b$. The spectral indices and corresponding spline fits obtained from EPOS-LHC and HIJING 1.3 simulations are shown in Figure \ref{fig-7}. As expected, large differences between the spectral indices of proton and iron distributions can be observed depending on the initial primary energy and the interaction model used. Since the differences between proton and iron spectra are larger than the model uncertainties and because muons will approximately have the same transverse momentum as their parent particles the difference between different primary mass compositions will be visible in the lateral separation distribution of muons. 

Using IceCube/IceTop coincident events it is possible to reconstruct double track events in IceCube as shown in Sec. \ref{sec-21} to obtain the lateral separation and to use IceTop energy reconstructions \cite{Ref25} to estimate the primary energy of the events. This allows measurement of the spectral index of the LS muon spectrum in bins of the primary energy and thus studies on cosmic ray mass composition based on high $p_\mathrm{T}$ muons. Moreover, dedicated simulations allow testing of different hadronic interaction models and pQCD predictions at high energies and low Bjorken-x.

\begin{figure}[t]
\includegraphics[width=0.48\textwidth,clip]{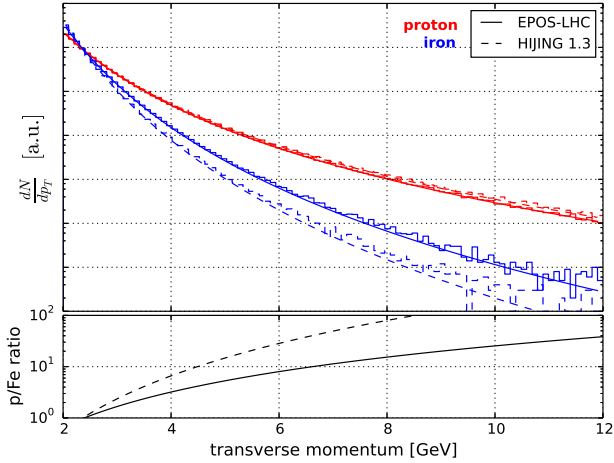}
\caption{Simulated $p_\mathrm{T}$ distributions for an incident energy spectrum of $dN/dE=E^{-\gamma}$ in the range $10^4\,\mathrm{GeV}\leq E_{\mathrm{prim}}\leq 10^8\,\mathrm{GeV}$ for EPOS and HIJING (upper panel). The parameters of the corresponding power law fits (lines) can be found in Table \ref{tab-1}. The lower panel shows the corresponding proton to iron ratio.}
\label{fig-8}      
\end{figure}

For an incident cosmic ray spectrum of the form $dN/dE\propto E^{-\gamma}$ where the spectral index $\gamma$ changes from $2.7$ to $3.0$ at $3\,\mathrm{PeV}$ (the knee) the transverse momentum for a certain primary energy is generated using the spectral index obtained from the spline fits in Figure \ref{fig-7}. The resulting $p_\mathrm{T}$ distributions from $10^7$ collisions generated with EPOS and HIJING for primary energies in the range of $10^4\,\mathrm{GeV}$ to $10^8\,\mathrm{GeV}$ are shown in Figure \ref{fig-8}. Following an $E^{-\gamma}$ primary spectrum $95\%$ of the events are generated with initial energies of $\log_{10}(E_{\mathrm{prim}}/\mathrm{GeV})<4.77$ using similar spectral indices (see Figure \ref{fig-7}). Thus, the resulting spectra can be approximated by power law distributions and are once again fitted using a power law with the best fit parameters shown in Table \ref{tab-1}. The iron distributions show a significantly steeper spectrum and at high $p_\mathrm{T}$ the expected rates for proton primaries are roughly ten times higher than for iron (lower panel).

Using this method we can simulate LS muon events by adding high $p_\mathrm{T}$ muons from pion, kaon, and heavy hadron decays to pre-simulated CORSIKA air showers. Transverse momentum and energy at generation are obtained from the parent particle spectrum according to the primary energy. Since the fraction of primary collisions producing at least one high $p_\mathrm{T}$ particle is known from the simulations, the flux of laterally separated muons can then be normalized to the known muon flux at the ground \cite{Ref9,Ref26}. After detector simulation, triggering, and filtering these simulations allow an optimization of the double track reconstruction for IC86 and the development of selection criteria for a mass composition analysis based on high $p_\mathrm{T}$ muons. 

An analysis of laterally separated muons using one year of IceCube data in its final 86-string configuration is in preparation where data from the in-ice detector is used to measure the lateral separation of muons and IceCube/IceTop coincident events allow studies on the cosmic ray mass composition. Furthermore, this analysis will include a dedicated search for triple track signatures introduced in Sec. \ref{sec-2}. Therefore we modified the reconstruction algorithm described in Sec. \ref{sec-21} based on a triple track hypothesis with a clustering into three sets of hits, where the largest of the clusters is defined as the bundle track and the other two sets of hits as LS muon tracks respectively. Although the flux will be low, this will extend high $p_\mathrm{T}$ muon analyses to very unique signal events with known origin that will therefore be a direct probe of the underlying primary cosmic ray interaction.


\section{Conclusions}
\label{sec-4}
Laterally separated muons can be used as a direct probe of high energy cosmic ray interactions because high energy particles with a high transverse momentum are typically produced in the first interaction. It was shown that the lateral separation of muons can be measured in IceCube using dedicated reconstruction methods and selection criteria. Data from the 59-string configuration indicates a transition in the lateral separation distribution from an exponential to a power law, as expected from pQCD. Moreover, simulations with most recent hadronic interaction models have shown that LS muons produced by the decay of high $p_\mathrm{T}$ particles are sensitive to the mass composition of the initial cosmic ray. Depending on the interaction model used the proton and iron spectra show significant differences. At high $p_\mathrm{T}$ the expected rates for proton primaries are roughly ten times higher than for iron which is larger than the uncertainties between different interaction models and will therefore be visible in the lateral separation distribution. Using IceCube/IceTop coincident events allows measurement of the primary mass composition as a function of the primary energy. Hence, the analysis of LS muons from cosmic ray air showers using data from IceCube/IceTop in its 86-string configuration will be sensitive to the primary mass composition and represents a complementary approach of composition analysis with respect to other cosmic ray experiments. Furthermore, studies on high $p_\mathrm{T}$ muons, together with dedicated simulations, provide useful information on high energy interactions of heavy nuclei at low Bjorken-x and can help to understand uncertainties in phenomenological models as well as test pQCD predictions.

\begin{table}[t]
\centering    
\begin{tabular}{c | c c c }
model & $a$ & $b$ & $\chi^2/DOF$ \\
\hline
\hline
{} & \multicolumn{3}{c}{proton} \\
\hline
EPOS-LHC & $1.1\cdot 10^8$ & $-5.6$ & $142/97$ \\
HIJING 1.3 & $9.7\cdot 10^7$ & $-5.4$ & $193/97$ \\
\hline
\hline
{} & \multicolumn{3}{c}{iron} \\
\hline
EPOS-LHC & $7.8\cdot 10^8$ & $-7.8$ & $575/97$ \\
HIJING 1.3 & $2.1\cdot 10^9$ & $-9.0$ & $1238/95$ \\
\end{tabular}
\caption{Best fit parameters of power law fits shown in Figure \ref{fig-7}. The estimated statistical errors are in the order of less than $1\%$.}
\label{tab-1}   
\end{table}


\end{document}